\documentclass[showpacs,12pt,preprintnumbers,amsmath,amssymb,floatfix]{revtex4-1}
\usepackage{graphicx}
\usepackage{caption}
\usepackage{subcaption}
\usepackage{rotating}
\usepackage{bm}

\newcommand{\subscript}[1]{\ensuremath{_{\textrm{#1}}}}

\begin{document}
\title{Prediction of A2 to B2 Phase Transition in the High Entropy Alloy MoNbTaW}
\author{William Paul Huhn}
\email{wph@andrew.cmu.edu}
\affiliation{Carnegie Mellon University, Pittsburgh, PA 15213}
\author{Michael Widom}
\affiliation{Carnegie Mellon University, Pittsburgh, PA 15213}
\date{June 21st, 2013}
\begin{abstract}
In this paper we show that an effective Hamiltonian fit with first principles 
calculations predicts an order/disorder transition occurs in the high entropy 
alloy MoNbTaW.  Using the Alloy Theoretic Automated Toolset, we find T=0K 
enthalpies of formation for all binaries containing Mo, Nb, Ta, and W, and in 
particular we find the stable structures for binaries at equiatomic concentrations 
are close in energy to the associated B2 structure, suggesting that at intermediate 
temperatures a B2 phase is stabilized in MoNbTaW. Our previously published 
hybrid Monte Carlo/molecular dynamics results for the MoNbTaW system are 
analyzed to identify certain preferred chemical bonding types. A mean field 
free energy model incorporating nearest neighbor bonds is derived, allowing 
us to predict the mechanism of the order/disorder transition.  We find the 
temperature evolution of the system is driven by strong Mo-Ta bonding.  
Comparison of the free energy model and our hybrid Monte Carlo/molecular dynamics
results suggest the existence of additional low-temperature phase transitions 
in the system likely ending with phase segregation into binary phases.
\end{abstract}
\maketitle
\section{Introduction}

Alloys of 4 or more constituent chemical species, each having 
roughly equal concentration, are known as ``high entropy alloys'' 
or HEAs.  Two properties justify research interest.  The first, 
relevant to material design, is that these alloys exhibit a 
``cocktail effect'' ~\cite{Kozak13}, whereby selection of individual 
constituent species tunes various material properties of the 
alloys.  The second, relevant to material growth and theoretical 
modeling, is that simple lattices are stabilized at moderate and 
high temperatures.  For a solid solution of N species, each of
concentration $1/N$, the high temperature entropy limit is 
$k_{B} {\rm ln} N$.  Since all atomic species play equivalent roles, 
BCC and FCC phases form rather than complex intermetallic phases
~\cite{Yeh04_1,Yeh04_2}.  Experimentally, many samples consist 
of a single simple phase, and those with multiple phase regions 
contain simple phases like A2 (BCC) and B2 (CsCl) rather than 
complex intermetallic structures.  This greatly simplifies 
theoretical modeling, as simple BCC and FCC models with ideal 
entropy may be used to understand the thermodynamic stability 
of these materials.

One research focus of HEAs are alloys containing the refractory metals 
Mo, Nb, Ta, and W.  These species are notable for their high melting 
temperatures, between 2750K and 3695K.  Alloying may be used to increased 
the melting temperature of HEAs, which has the added effect of increasing 
the onset temperature of alloy softening, which occurs  around 50\% to 60\% of melting temperature.  
Additionally, atomic radius mismatch creates localized distortions in the 
lattice, inducing solid solution strengthening~\cite{Yeh04_1,Zhou07}.  
High melting temperatures and strong mechanical properties make these HEAs useful 
for aerospace and other applications.  Previous experimental work in MoNbTaW and MoNbTaVW 
~\cite{Senkov10,Senkov11} show single-phase BCC structures with hardnesses 
of 4.5 and 5.3 GPa and high yield strength up to 1873 K, however they are 
brittle at room temperature.  These high melting temperatures 
combined with the expected phase stability over large temperature ranges 
makes experimental study of the thermodynamic properties of these alloys 
problematic, as it is unlikely that thermodynamic equilibrium can be 
achieved on experimental time scales.  Accordingly, only high temperature 
(\textgreater2000K) phase diagrams exist for the refractory metal binaries, 
all of which present a single A2 phase across the entire composition range, 
with liquidus and solidus lines nearly coincident.  Due to lack of equilibration, 
what experimental evidence of thermodynamic properties at low temperatures 
exists is of doubtful accuracy.  To study phase stability of such alloys, 
first principles calculations are the most accurate method currently available.  

We are here interested in the possibility of the A2 phase transitioning 
to B2 as temperature drops.  In a previous paper ~\cite{Widom13}, we 
developed a hybrid Monte Carlo/molecular dynamics (MC/MD) method that 
produced such a phase transition.  Based upon similar atomic radius 
and electronegativity, we found that MoNbTaW orders as a pseudobinary 
system consisting of Group 5 (Ta,Nb) and Group 6 (W,Mo) atoms, with intergroup 
bonding stronger than intragroup.  In this paper, we propose an effective
Hamiltonian that exhibits an A2 $\rightarrow$ B2 transition, while
 showing deviations from this pseudobinary model.

\section{T=0K Binary Enthalpies of Formation}
To generate ground state structures, we use the Alloy Theoretic Automated 
Toolkit (ATAT) ~\cite{vanDeWalle02_1,vanDeWalle02_2,vanDeWalle02_3,vanDeWalle09}, 
a framework that iteratively generates a cluster expansion, based on 
ab-initio total energies, to suggest new candidate ground state structures 
for a given lattice type.  We applied ATAT to all six binary combinations 
of Mo, Nb, Ta, and W on a BCC lattice. Between 60 and 100 structures were generated per 
binary.  The K-Point Per Reciprocal Atom (KPPRA) density was fixed to 9000 
for all binaries.  All binaries finished with crossvalidation scores less 
than 6.3 meV/atom.  Subsequent relaxation and convergence of predicted ground state and metastable 
structures in k-point density was then performed.  To calculate total energies, 
we use VASP (the Vienna Ab-Initio Simulation Package) 
~\cite{KresseHafner93,KresseFurthmuller96}, a plane wave ab-initio package 
implementing PAW pseudopotentials ~\cite{Blochl94}.  We use the PBE density 
functional ~\cite{PerdewBurkeErnzerhof96}, with default energy cutoffs for 
total energy calculations.  Relaxation was performed at P=0.  To our knowledge, 
no examinations of refractory metal binaries have been performed using this 
functional.  

We identify structures by the notation [chemical formula].[Pearson symbol], 
using only the Pearson symbol when the chemical formula is implicitly 
understood.  Common candidate structures for these binaries include cP2 
at 50\% composition (Strukturbericht B2, prototype CsCl), oA12 at 50\% 
composition (Strukturbericht B2\subscript{3}), tI6 at 33.3\% and 66.7\% 
composition (Strukturbericht C11\subscript{b}, prototype MoSi\subscript{2}), 
and cF16 at 25\% and 75\% composition (Strukturbericht D0\subscript{3}, 
prototype BiF\subscript{3}).  oA12, a variant of cP2 with antiphase 
boundaries, is of special importance as it has been identified by previous 
work ~\cite{Hart05,Blum05} using cluster expansion methods as a potential 
ground state structure in Mo-Nb, Nb-W, and Ta-W.
 
Our results for binaries may be summarized as follows.  Mo-Ta shows 
strongest bonding, with strongest enthalpy of formation of -186 meV/atom.  
Mo-Nb and Ta-W have nearly equal bonding at -103 meV/atom and -110 meV/atom 
respectively, with Nb-W the weakest of the intergroup binaries at 
-53 meV/atom. The intragroup binaries Mo-W and Nb-Ta are essentially 
ideal (i.e. vanishing enthalpy).   This ordering is consistent with 
electronegativity and atomic radius differences~\cite{Widom13}.  Mo-Ta and 
Mo-Nb have roughly symmetric convex hulls about equiatomic 
concentration, whereas Nb-W and Ta-W are biased towards high W 
concentrations with minima at 66.7\% W.  Both sets of observations 
are consistent with experimental observations that intergroup alloys
Mo-Ta, Mo-Nb, Nb-W, and Ta-W exhibit cleavage near equiatomic 
concentrations, while Mo-W and Nb-Ta exhibit flow~\cite{vanTorne66}.  
We confirm that in these systems cP2 is not the stable structure 
at equiatomic concentration, with the exception of Mo-W where 
enthalpies of formation deviate negligibly from ideality, and that 
oA12 is stable for many of the binaries.  However, with the 
exception of Nb-W, cP2 is within 10 meV/atom of the convex hull.  
This suggests that a B2 phase could be stabilized at intermediate 
temperature through the entropy of intragroup mixing.  Below we compare
our predictions for individual binaries with prior literature.

\begin{figure}
	\begin{subfigure}[b]{0.49\textwidth}
		\includegraphics[trim = 0mm 0mm 0mm -10mm, clip, width=\textwidth]{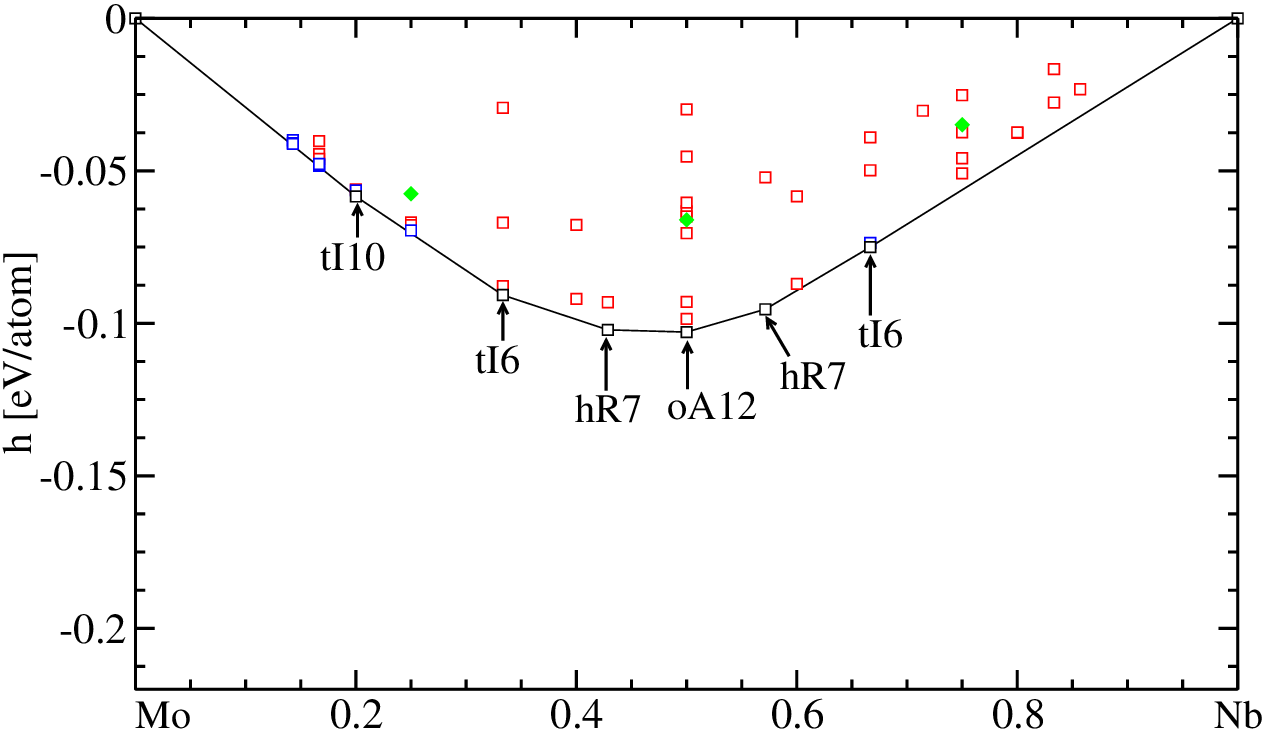}
		\caption{Mo-Nb}
	\end{subfigure}
	\begin{subfigure}[b]{0.49\textwidth}
		\includegraphics[trim = 0mm 0mm 0mm -10mm, clip, width=\textwidth]{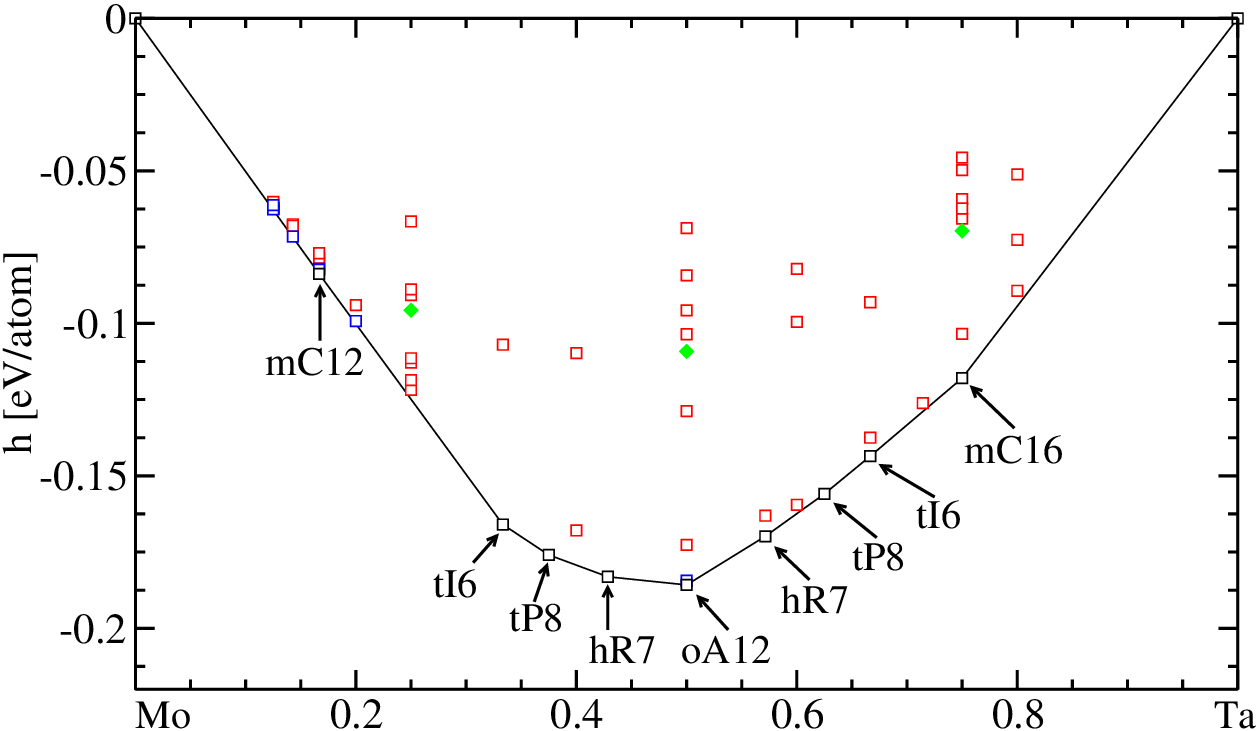}
		\caption{Mo-Ta}
	\end{subfigure}
	\begin{subfigure}[b]{0.49\textwidth}
		\includegraphics[trim = 0mm 0mm 0mm -10mm, clip, width=\textwidth]{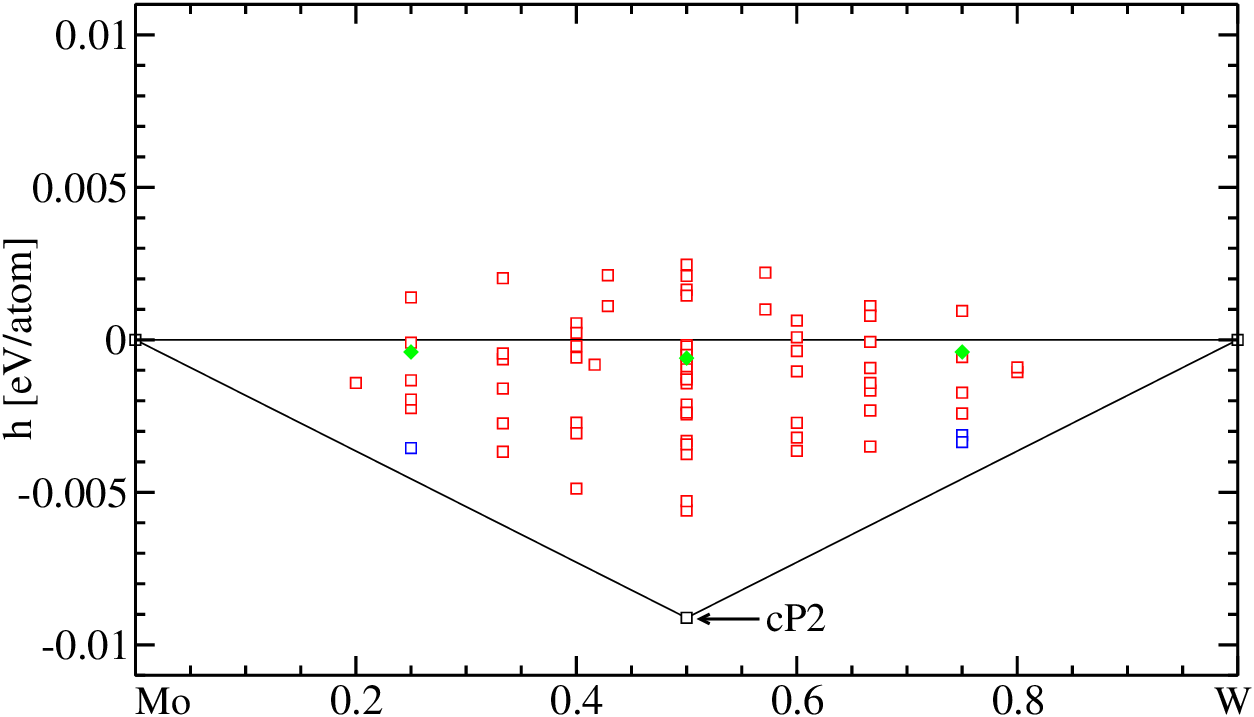}
		\caption{Mo-W}
	\end{subfigure}
	\begin{subfigure}[b]{0.49\textwidth}
		\includegraphics[trim = 0mm 0mm 0mm -10mm, clip, width=\textwidth]{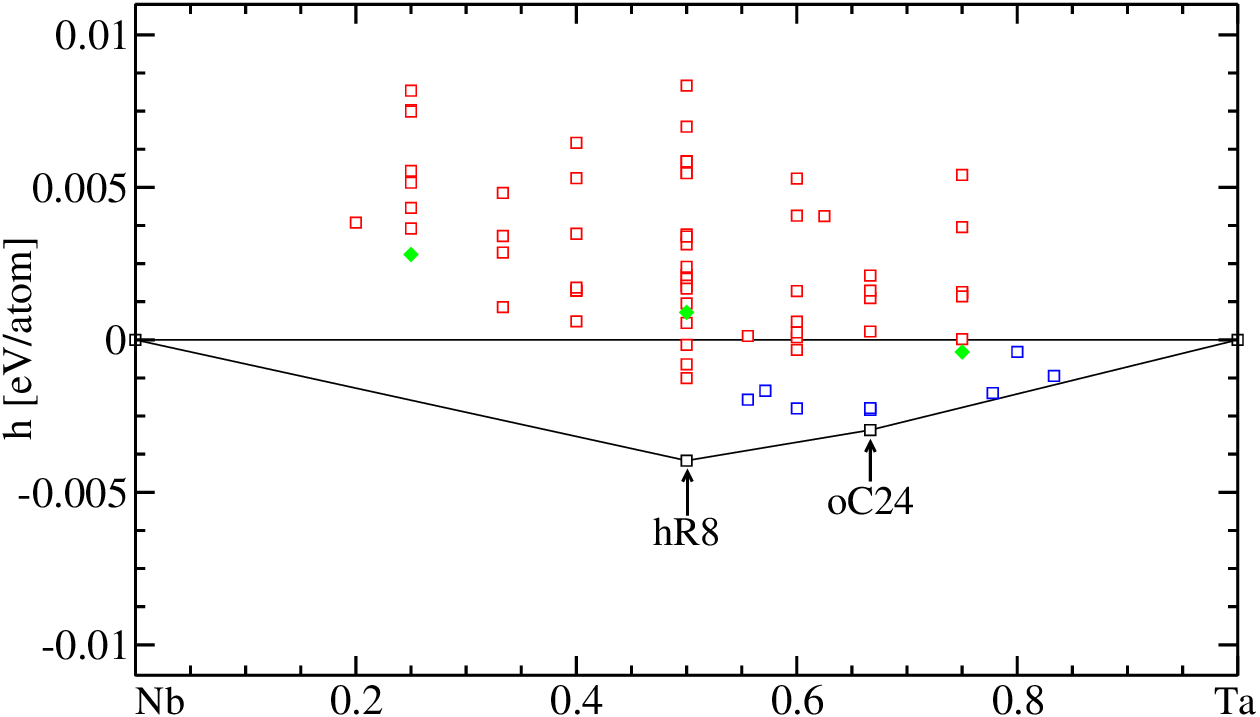}
		\caption{Nb-Ta}
	\end{subfigure}
	\begin{subfigure}[b]{0.49\textwidth}
		\includegraphics[trim = 0mm 0mm 0mm -10mm, clip, width=\textwidth]{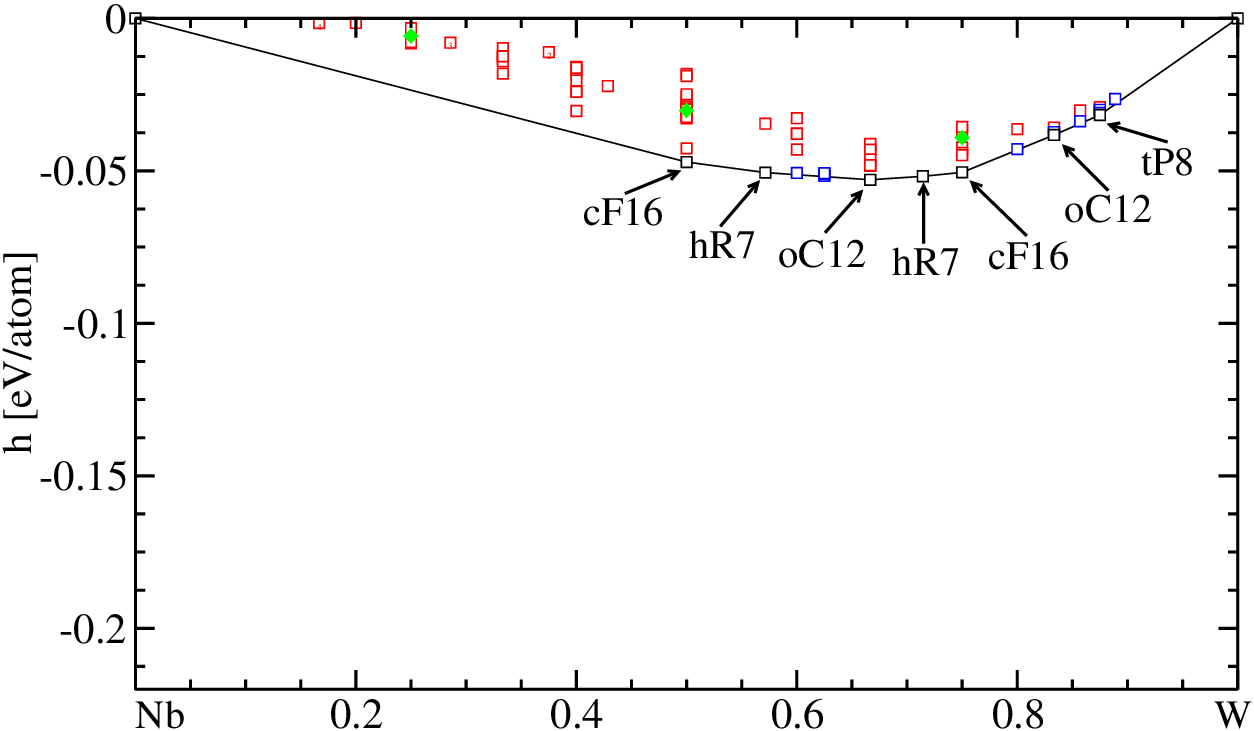}
		\caption{Nb-W}
	\end{subfigure}
	\begin{subfigure}[b]{0.49\textwidth}
		\includegraphics[trim = 0mm 0mm 0mm -10mm, clip, width=\textwidth]{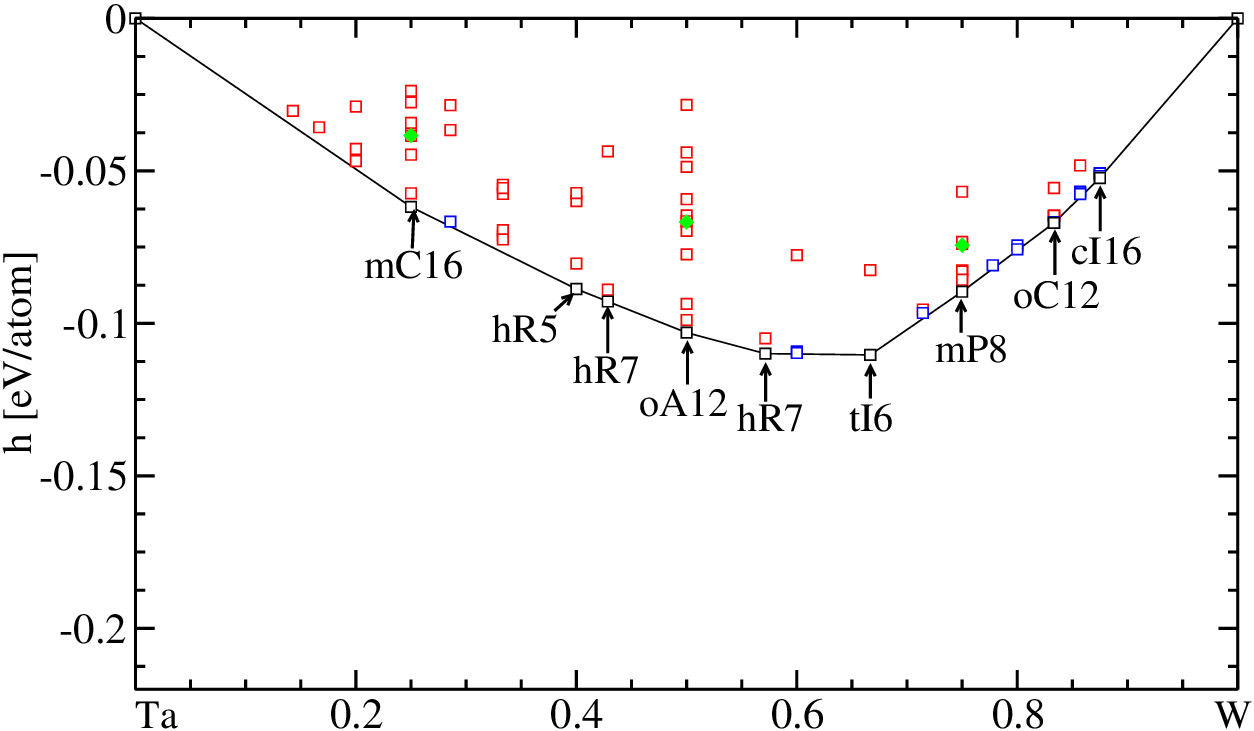}
		\caption{Ta-W}
	\end{subfigure}
	\caption{First principles enthalpies of formation for binaries.  Black squares denotes structures on the convex hull, blue squares denotes structures within 2 meV/atom of the convex hull, and red squares above 2 meV/atom.  Filled green diamonds denote 16 atom BCC SQS (special quasirandom  structures)~\cite{Jiang04}.  The scales of Mo-W and Nb-Ta differ from the scales from the other four binaries.}
	\label{fig:EOF}
\end{figure}

\subsection{Mo-Nb}
Previous first principles calculations on this system have been 
performed by Curtarolo {\it et al.}~\cite{Curtarolo08} and Blum and 
Zunger~\cite{Blum05}.  While both observe the cP2, Mo-rich 
cF16, and Mo-rich and Nb-rich tI6 structures stable using the 
LDA functional, Blum and Zunger contend that their mixed-basis
cluster expansion (MBCE) fit indicates that the cP2 and Nb-rich 
tI6 are not stable and that at equiatomic concentration oA12 is
stable (although LDA predicts this structure to be unstable).  
Our PBE study predicts oA12 is stable and minimizes 
the enthalpy of formation at -102.9 meV/atom, while cP2 lies 
4.3 meV/atom above.  We find both Mo-rich and Nb-rich tI6 lie 
on the convex hull.  Mo-rich cF16 lies 1.0 meV/atom off the convex hull, 
well within margin of error, but Ta-rich cF16 lies substantially above 
at 18.9 meV/atom.  The Mo-rich side of the convex hull is more detailed
than the Nb-rich side, though the convex hull is generally symmetric about equiatomic 
concentration.  
 
\subsection{Mo-Ta}
Previous first principles work on this system has been performed on 
this system by Blum and Zunger~\cite{Blum04,Blum05} and Turchi {\it et 
al.}~\cite{Turchi04}.  While Blum and Zunger observe the cP2 and 
Mo-rich and Ta-rich tI6 to be stable using the LDA functional, their
MBCE states that the Ta-rich tI6 is not stable.  Turchi {\it et al.}, using 
a cluster variation method approach combined with first principles 
calculations, find an A2 to B2 transition at 1772K and 47\% Ta.  
oA12 minimizes enthalpy of formation for the system at -185.8 
meV/atom, with cP2 1.4 meV/atom above.  We find Mo-rich and Ta-rich tI6 
to both be stable, in agreement with LDA results but not Blum and 
Zunger's MBCE.  van Torne and Thomas ~\cite{vanTorne66} observed 
non-ideal behavior in Mo-Ta BCC solid solutions at T=273K, in line 
with the tendency towards chemical ordering, though they attribute 
this to concentration gradients in their sample which were likely not 
in equilibrium.

\subsection{Mo-W and Nb-Ta}
For binaries Mo-W and Nb-Ta, at 2000K and 2600K, respectively, 
experimental enthalpies of mixing are low for all compositions 
and activities perfectly follow Raoult's Law, suggesting an ideal 
solution with no chemical bonding.  This is in agreement 
with our first principles calculations, which have a minimum enthalpy 
of formation of -10 meV/atom and -3 meV/atom for Mo-W and Nb-Ta, 
respectively,  indicating nearly perfect mixing of atomic species.  
This is in agreement with Villar's empirical criteria~\cite{Villars83}, 
as both binaries consist of BCC metals with similar electronegativity 
and atomic radius. To our knowledge, no other first principles results 
for these binaries exist in the literature.  

\subsection{Nb-W}
For Nb-W significant deviations from symmetry in the convex hull are 
observed, with no stable structures found on the Nb-rich side of the 
convex hull.  The equiatomic concentration stable structure is cF16
(prototype NaTl), with cP2 28.2 meV/atom above.  However, the minimum 
enthalpy of formation structure is an oC12 structure at 66.7\% W with 
enthalpy of formation at -52.9 meV/atom. Blum and Zunger ~\cite{Blum05} 
also found a detailed W-rich side of the convex hull, and a Nb-rich side
containing only one structure that negligibly affects the shape of the
convex hull.   In particular, our PBE predicts W-rich tI6 is unstable by 6.6 
meV/atom, which does not agree with LDA results, but does agree with 
Blum and Zunger's MBCE.

\subsection{Ta-W}
Of all binaries considered, Ta-W is the most studied.  Experimental work
~\cite{Singhal73} indicates negative deviation from Vegard's law for 
lattice constants, asymmetry of the enthalpy of mixing towards the 
Ta-rich side, and significant deviation from ideality of activities at 
1200K, suggesting the presence of short range order.  Early work by
Turchi {\it et al.} ~\cite{Turchi01} focused only on cP2 and cF16, 
both Ta- and W-rich, finding all three structures stable.   Order-disorder 
transitions were studied by Masuda-Jindo {\it et al.}~\cite{MasudaJindo08} 
using a cluster expansion fitted from first principles calculations on 
random alloys, finding a second order A2 to B2 phase transition first 
appearing around 1000K near equiatomic concentration.   We find oA12 
to be stable at equiatomic concentration at an enthalpy of formation 
of -103.1 meV/atom, with cP2 9.4 meV/atom above, but W-rich tI6 minimizes 
enthalpy of formation at -110.3 meV/atom.  That the convex hull leans W-rich 
is supported by the cluster expansions of Blum and Zunger {\it et al.}~\cite{Blum05} and 
Hart {\it et al.}~\cite{Hart05}, both of whose cluster expansions have W-rich 
tI6 on the convex hull and nearly minimizing the enthalpy of formation.  While 
the convex hull leaning W-rich disagrees with experimental evidence, this is 
in line with other theoretical results and at 1200K it is unlikely experimental 
results can be properly equilibrated.  

\section{Quaternary and MC/MD Results}
By the third law of thermodynamics, at T=0K we expect MoNbTaW to 
chemically order or to phase segregate into well-ordered structures.  
Considering only binary structures, we find at the equiatomic 
concentration for MoNbTaW the stable coexisting phases are MoTa.oA12, 
NbW.cF16, TaW\subscript{2}.tI6, and Mo\subscript{3}Nb\subscript{4}.hR7, 
with an average enthalpy of formation of -117 meV/atom.  That the first 
3 are stabilized in the quaternary is not surprising as they are the enthalpy 
minimizing structures in their respective binaries.  MoTa.oA12 in particular 
is stabilized as it has enthalpy of formation a factor of 2 or more larger than all other 
binaries.  While MoNb.oA12 is the enthalpy of formation minimizing structure 
for Mo-Nb, the convex hull for the Mo-Nb system is shallow near equiatomic 
concentration and thus little enthalpy is lost segregating out 
Mo\subscript{3}Nb\subscript{4}.hR7 instead.  This allows TaW\subscript{2}.tI6 
and NbW.cF16 to be stabilized while maintaining the proper stoichiometry.  

\begin{figure}
	\includegraphics[trim = 0mm 0mm 0mm 0mm, clip, width=0.75\textwidth]{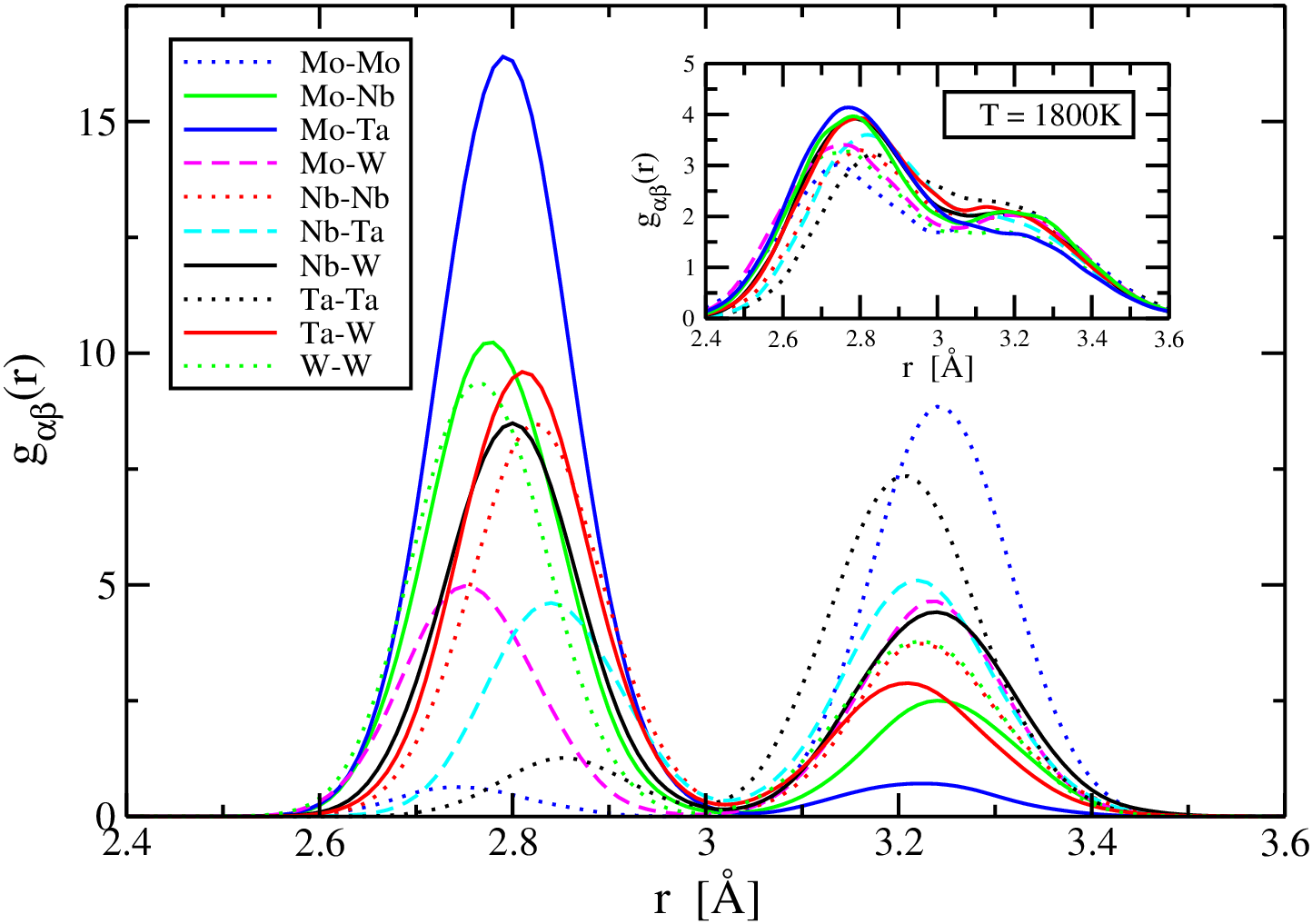}
	\caption{Partial density functions obtained from MC/MD for the MoNbTaW BCC phase, showing first and second nearest neighbors.  Solid lines denote PDFs between species from differing groups, dashed lines between differing species from the same group, and dotted lines between the same species.  The main figure shows results for T=300K, and the inset for T=1800K}
	\label{fig:pdf}
\end{figure}

We here examine our previously reported MC/MD calculations~\cite{Widom13} 
in MoNbTaW in light of the T=0K binary phase diagrams.  As 
previously reported and reproduced here, the nearest neighbor partial 
distribution functions (NN-PDF) at $T=300K$ in Figure \ref{fig:pdf} 
generally agree with the magnitudes of enthalpies obtained from the 
previous section, though deviations exist.  Mo-Ta has the largest 
enthalpy of formation, and indeed the NN-PDF for Mo-Ta shows the 
strongest peak, suggesting enthalpic stabilization.  Next comes Nb-Mo 
and Ta-W, which are nearly degenerate in enthalpy at equiatomic
composition, and their NN-PDF are nearly equal.  Nb-Nb and W-W come next, which anomalously 
deviate from pseudobinary behavior, as they are intragroup peaks.  
Nb-W is the weakest of the intergroup bonds and has the smallest NN peak, though it is still 
significant.  The remainder are all intragroup, with the NN peaks 
for Mo-Mo and Ta-Ta very nearly zero.  As a possible 
explanations for the presence of W-W and Nb-Nb NN bonds, note that 
at $T=0K$ two of the stable structures near this stoichiometry, 
TaW\subscript{2}.tI6 and Mo\subscript{3}Nb\subscript{4}.hR7, are 
respectively W-rich and Nb-rich, yielding W-W and Nb-Nb bonds.  The anomalous 
NN peaks may indicate the tendency of the system to undergo partial phase 
segregation at this temperature.  The next-nearest neighbor 
partial distribution functions (NNN-PDF) at $T=300K$ show exactly 
opposite ordering, with Mo-Mo and Ta-Ta showing strong peaks and Mo-Ta 
essentially zero, characteristic of cP2-like ordering.  

For $T=1800K$, shown in the inset of Figure \ref{fig:pdf}, all NN peaks 
have similar magnitude, as do NNN peaks.  By $T=1800K$, ordering has 
apparently been lost and the system is a random BCC solid solution.  We propose 
the following temperature evolution.  At high temperature, the system 
is a nearly ideal disordered BCC solution, as shown by the $T=1800K$ 
PDF.  As the temperature of the system decreases, the system undergoes 
an order/disorder phase transition, most likely to cP2-like alternation 
with the pattern of chemical order reflecting preferred bonding types 
amongst species.  As the temperature further decreases, additional phase 
transitions are possible, culminating in phase segregation into the $T=0K$ 
coexisting phases. 
 
\section{A Free Energy Model for MoNbTaW}
To calculate qualitative details of the order-disorder transition in 
MoNbTaW, we derive a mean field free energy model for chemical ordering 
within the BCC phase.   Previous work on phase stability using 
Monte Carlo methods has been performed by del Grosso {\it et al.}
~\cite{delGrosso12_1,delGrosso12_2}, however their empirical 
interaction model predicts Mo and Ta avoid NN bonding at T=0K.  This yields 
different low-temperature phase segregation behavior than what first 
principles unambiguously predicts.  As we are using a mean field 
model, local elastic distortion of the crystal lattice due to atomic radius 
mismatch, a known strong effect in HEAs~\cite{Zhang08}, is incorporated in 
our study only in so far as it affects the binary enthalpies of formation.

We work in the Gibbs ensemble at fixed temperature, pressure, and chemical 
composition in a system with $N$ atoms and $d$ chemical species, all of 
which have the same number of atoms $N/d$.  We consider only nearest neighbor 
interactions, which for a BCC lattice exist between cell center and cell 
vertex sites, yielding an enthalpy in the form of an effective Hamiltonian
\begin{equation}
H=\sum_{<ij>}\sum_{\alpha\beta}\sigma_{\alpha}(i)b_{\alpha\beta}\sigma_{\beta}(j),
\label{eq:NN_H}
\end{equation}
where $i$ and $j$ denotes all possible sites, summation over $<ij>$ denotes 
summation over all nearest neighbor bonds, summation over $\alpha$ and 
$\beta$ denotes summation over all possible species, $b_{\alpha\beta}$ 
denotes the nearest neighbor bond strength between species $\alpha$ and 
$\beta$, and $\sigma_{\alpha}(i)$ is $1$ if site $i$ contains species 
$\alpha$ and 0 otherwise. 

Anticipating cP2-like ordering, we now define the quantities
\begin{equation}
e_{\alpha} = \frac{2}{N}\sum_{i~\in~vertices}\sigma_{\alpha}(i)
\label{eq:e_a}
\end{equation}
and
\begin{equation}
o_{\alpha} = \frac{2}{N}\sum_{i~\in~centers}\sigma_{\alpha}(i),
\label{eq:o_a}
\end{equation}
which are, respectively, the concentration of species $\alpha$ on cell vertex 
(``even'') sites and the concentration of species $\alpha$ on cell center 
(``odd'') sites.  Inverting Eq. (\ref{eq:e_a}) and (\ref{eq:o_a}), we rewrite
$\sigma_{\alpha}(i)$ as
\begin{equation}
\sigma_{\alpha}(i) = (e_{\alpha},o_{\alpha}) + \delta \sigma_{\alpha}(i),
\label{eq:sigma}
\end{equation}
where the first term is $e_{\alpha}$ if $i$ is a cell vertex and 
$o_{\alpha}$ if $i$ is a cell center.  As all NN bonds are between cell 
centers and vertices, we may rewrite Eq. (\ref{eq:NN_H}) as
\begin{equation}
H=\sum_{i~\in~centers}\sum_{j~\in~NN(i)}\sum_{\alpha\beta}(o_{\alpha} + \delta \sigma_{\alpha}(i))b_{\alpha\beta}(e_{\beta}+\delta \sigma_{\beta}(j)).
\label{eq:NN_H_2}
\end{equation}
In a mean field approximation, terms in $\delta \sigma$ vanish.  The 
remaining term is independent of $i$ and $j$ and thus scales as the total 
number of bonds $4N$, giving an enthalpy per atom of:
\begin{equation}
h=H/N=\sum_{\alpha\beta}o_{\alpha}\Omega_{\alpha\beta}e_{\beta}
\label{eq:h}
\end{equation}
where $\Omega_{\alpha\beta}=4b_{\alpha\beta}$.

We now introduce the ideal entropy approximation, assigning an entropy per atom
of
\begin{equation}
s = -\frac{k_{B}}{2}\sum_{\alpha}(e_{\alpha}ln(e_{\alpha})+o_{\alpha}ln(o_{\alpha})).
\label{eq:entropy_per_atom}
\end{equation}
As $e_{\alpha}$ and $o_{\alpha}$ are bounded between 0 and $2/d$ and sum to 1, 
this entropy can only vanish in the case where $d=2$, whereas we will apply 
this free energy to HEAs with $d \ge 4$.  Hence this entropy violates the third 
law of thermodynamics were we to na\"{i}vely extrapolate it to T=0K.  This 
is a natural consequence of B2 ordering, with 2 unique sites occupied by $d > 2$ 
chemical species, requiring disorder on the sites and creating entropy of mixing.  
To predict phase transitions at lower temperatures, both a unit cell with number 
of unique sites divisible by 4, and more interaction terms, must be included.  
Including the entropic contribution $-Ts$, we obtain a free energy per atom of
\begin{equation}
g = \sum_{\alpha\beta}o_{\alpha}\Omega_{\alpha\beta}e_{\beta}+\frac{k_{B}T}{2}\sum_{\alpha}(e_{\alpha}ln(e_{\alpha})+o_{\alpha}ln(o_{\alpha})).
\label{eq:dg_full}
\end{equation}

We now examine ordering tendencies driven by the energetics of the 
system.  Suppose we have an enthalpy of general quadratic form
\begin{equation}
h = {\bm \psi^{T}} \Omega {\bm \psi}
\label{eq:H_mat1}
\end{equation}
where ${\bm \psi}$ is a vector containing $n+r$ variables, where $n$ of
the variables are independent and $r$ are dependent, and $\Omega$ 
is an $(n+r) \times (n+r)$ dimensional symmetric matrix.  We rewrite this enthalpy 
in terms of only independent variables.  Define ${\bm \psi}$ 
with the first $n$ entries independent, so that it may be decomposed into 
an $n$-dimensional vector ${\bm \psi_{i}}$ containing independent variables 
and an $r$-dimensional vector ${\bm \psi_{d}}$ containing dependent variables.  
Rewrite Eq. (\ref{eq:H_mat1}) in block diagonal form
\begin{equation}
h =
\left[
\begin{array}{cc}
{\bm \psi_{i}^{T}} & {\bm \psi_{d}^{T}} \\
\end{array}\right]
\left[
\begin{array}{cc}
\Omega^{ii} & \Omega^{id} \\
(\Omega^{id})^{T} & \Omega^{dd} \\
\end{array}\right]
\left[
\begin{array}{c}
{\bm \psi_{i}} \\
{\bm \psi_{d}} \\ 
\end{array}\right] \\
\label{eq:H_mat2}
\end{equation}
where $\Omega^{ii}$ is an $n \times n$ dimensional matrix, $\Omega^{id}$ 
an $n \times r$ matrix, and $\Omega^{dd}$ and $r \times r$ matrix.  Assume 
that ${\bm \psi_{d}}$ depends linearly on ${\bm \psi_{i}}$, so the 
constraints on the system may be written in form
\begin{equation}
{\bm \psi_{d}} = {\bm k} + D{\bm \psi_{i}}
\end{equation}
with ${\bm k}$ an $r$-dimensional vector and $D$ an $r \times n$ dimensional matrix.
Imposing this on Eq. (\ref{eq:H_mat2}) yields
\begin{equation}
h = {\bm \psi_{i}^{T}}\theta{\bm \psi_{i}} + B{\bm \psi_{i}}+C
\label{eq:H_mat3}
\end{equation}
where $\theta = \Omega^{ii}+\Omega^{id}D+D^{T}(\Omega^{id})^{T}+D^{T}\Omega^{dd}D$ 
is an $n \times n$ matrix, $B= 2{\bm k^{T}}((\Omega^{id})^{T}+\Omega^{dd}D)$ a $n$-dimensional 
row vector, and $C={\bm k^{T}}\Omega^{dd}{\bm k}$.  If $\theta$ is invertible, $h$ has a 
unique extremum at
 \begin{equation}
{\bm \psi_{0}}=-\frac{1}{2}\theta^{-1}B^{T},
\end{equation} 
so Eq. (\ref{eq:H_mat3}) may be rewritten with a coordinate redefinition of 
${\bm \psi'} \equiv {\bm \psi} - {\bm \psi_{0}}$ as
\begin{equation}
h={\bm \psi'^{T}}\theta{\bm \psi'}+C-{\bm \psi_{0}^{T}}\theta{\bm \psi_{0}}.
\end{equation}
As $\Omega^{ii}$ and $\Omega^{dd}$ are symmetric, $\theta$ must 
be a real symmetric matrix and is therefore diagonalizable with 
orthogonal eigenvectors.  Its eigenvalues and eigenvectors yield 
information about preferential ordering in the system.

Returning to the special case of cP2 symmetry, there are $d$ 
constraints of the form
\begin{equation}
e_{\alpha} + o_{\alpha} = \frac{2}{d},
\label{eq:spec_occ}
\end{equation}
one for each $\alpha$, and two constraints imposing that each site class must be 
completely occupied:
\begin{equation}
\sum_{\alpha}{e_{\alpha}}=\sum_{\alpha}{o_{\alpha}}=1
\label{eq:site_con}
\end{equation}
However the two constraints of Eq. (\ref{eq:site_con}) are redundant, as required by
Eq. (\ref{eq:spec_occ}).  It follows that only $d-1$ of the original $2d$ variables 
are independent.  Here we take all $o_{\alpha}$ and one of the $e_{\alpha}$ to be 
dependent.  For cP2 ordering, translational symmetry and equiatomic composition 
require all components of ${\bm \psi_{0}}$ equal 1/d.  This yields an enthalpy of 
the form
\begin{equation}
h = \sum_{\alpha'\beta'}(e_{\alpha'}-\frac{1}{d})\theta_{\alpha'\beta'}(e_{\beta'}-\frac{1}{d})+\frac{1}{d^{2}}\sum_{\alpha\beta}\Omega_{\alpha\beta}
\label{eq:dg_final}
\end{equation}
where primed indices denote summation over independent species and unprimed 
indices denote summation over all species.  The enthalpy is invariant under 
the body centering operation $e_{\alpha} \rightarrow 2/d - e_{\alpha}$.  
This enthalpy has the natural form expected for an order-disorder transition.  
In particular, for the binary (d=2) system A-B, $h = \Omega_{AB}(e_{A}-\frac{1}{2})^{2}$ 
up to a constant, with $e_{A}-\frac{1}{2}$ the well-known order parameter 
for the order/disorder phase transition in the cP2 structure.  

We diagonalize $\theta$, obtaining its eigenvectors ${\bm v_{i}}$ 
with associated eigenvalues $\lambda_{i}$.  Any point in composition 
space may be written as ${\bm \psi'} = \sum_{i}\chi_{i}{\bm v_{i}}$, where 
$\chi_{i}$ is the associated normal coordinate for eigenvector ${\bm v_{i}}$.
Each normal coordinate is a linear combination of $e_{\alpha}$'s, with 
$\chi_{i} = 0~\forall~i$ corresponding to $e_{\alpha}=\frac{1}{d}~\forall~\alpha$.  
Normalizing the eigenvectors with a squared magnitude of unity, the enthalpy
becomes
\begin{equation}
h=\sum_{i} \lambda_{i}\chi_{i}^{2} + \frac{1}{d^{2}}\sum_{\alpha\beta}\Omega_{\alpha\beta}
\label{eq:H_diag}
\end{equation}
There are two cases for the temperature evolution of the system, depending 
on the spectrum of $\theta$. In the first case, $\theta$ has only 
positive or zero eigenvalues.  In this case, the enthalpy-minimizing 
configuration is $\chi_{i}=0~\forall~i$.  As this is also the entropy 
maximizing configuration, the system remains disordered at 
$e_{\alpha}=1/d~\forall~T$ .  No order/disorder phase transition occurs 
if $\theta$ has only non-negative eigenvalues.

In the second case, $\theta$ has negative eigenvalues leading to 
solutions with enthalpy less than the disordered solution.  Were 
it not for the constraint $ 0 \le e_{\alpha} \le 2/d$, the enthalpy 
would be unbounded below.  The $T=0K$ enthalpy-minimizing solution 
must lie on the boundary of configuration space, i.e. $e_{\alpha} = 0$ or 
$e_{\alpha} = 2/d$ (which are physically equivalent due to body centering 
symmetry) for at least one species $\alpha$, corresponding to at least one 
species showing perfect ordering at $T=0K$.  At any point in configuration 
space not on the boundary, we may always increase the normal coordinate 
corresponding to any negative eigenvalue to lower the enthalpy 
until the boundary is reached.  In general, multiple normal coordinates  
contribute to the solution.  Non-zero normal coordinates for positive eigenvalues 
may exist, as increasing the normal coordinates for positive eigenvalues may 
move the system in configuration space away from the boundary.  Subsequently 
increasing the normal coordinate for negative eigenvectors moves the system 
in configuration space back to the boundary, possibly with lower enthalpy than 
before.  As the enthalpy is independent of temperature, there must be a finite 
temperature where the entropic contribution to free energy (which grows 
proportionate to $T$) dominates the enthalpic contribution.  An order/disorder 
phase transition must occur if $\theta$ has at least one negative eigenvalue.

\begin{table}
\begin{tabular}{c|cc|cc}
   &\multicolumn{2}{c|}{Group 5}&\multicolumn{2}{c}{Group 6}\\
   &  Ta        & Nb        & W         & Mo \\
\hline
Ta &    0 &   1 & -34 & -53 \\
Nb &    1 &   0 & -14 & -30 \\
\hline
W  &  -34 & -14 &   0 &  -1 \\
Mo &  -53 & -30 &  -1 &   0 \\
\end{tabular}
\caption{Nearest neighbor bond strengths $b_{\alpha\beta}$ used in modeling the MoNbTaW BCC phase, in units of meV/atom.}
\label{tab:NNBondNum} 
\end{table}

\section{Results of Free Energy Model}

To determine if an order-disorder phase transition exists in MoNbTaW, we 
compute the NN bond strengths $b_{\alpha\beta}$.  
Using the structures previously obtained from ATAT, we reran the cluster expansion, 
restricting it to a single 2 body term, which yields the NN bond strengths 
between differing species given in Table \ref{tab:NNBondNum}.  Shown in 
Table \ref{tab:eigenvecval} are the eigenvalues of the $\theta$ matrix, 
the eigenvectors, and the associated dependent composition 
$e_{Mo}-1/d = \sum_{\alpha'}(e_{\alpha'}-1/d)$.  Positive signs for 
eigenvector components denote ordering on cell vertices and negative 
signs denote ordering on cell centers (only the relative sign between 
components are relevant).  Of the three possible modes, one has a 
negative eigenvalue, so an order/disorder phase transition must exist.  
The lowest enthalpy mode ordering the system ($\lambda$ = -701 meV/atom) 
indeed has strong ordering with opposite signs for Mo and Ta, supporting 
the assertion that Mo-Ta NN bonds drive the ordering of the system.  This 
mode also has strong ordering with opposite signs on Mo and Nb, in 
agreement with Mo-Nb as the second strongest bonding type.

\begin{figure}
	\includegraphics[trim = 0mm 0mm 0mm 0mm, clip, width=0.75\textwidth]{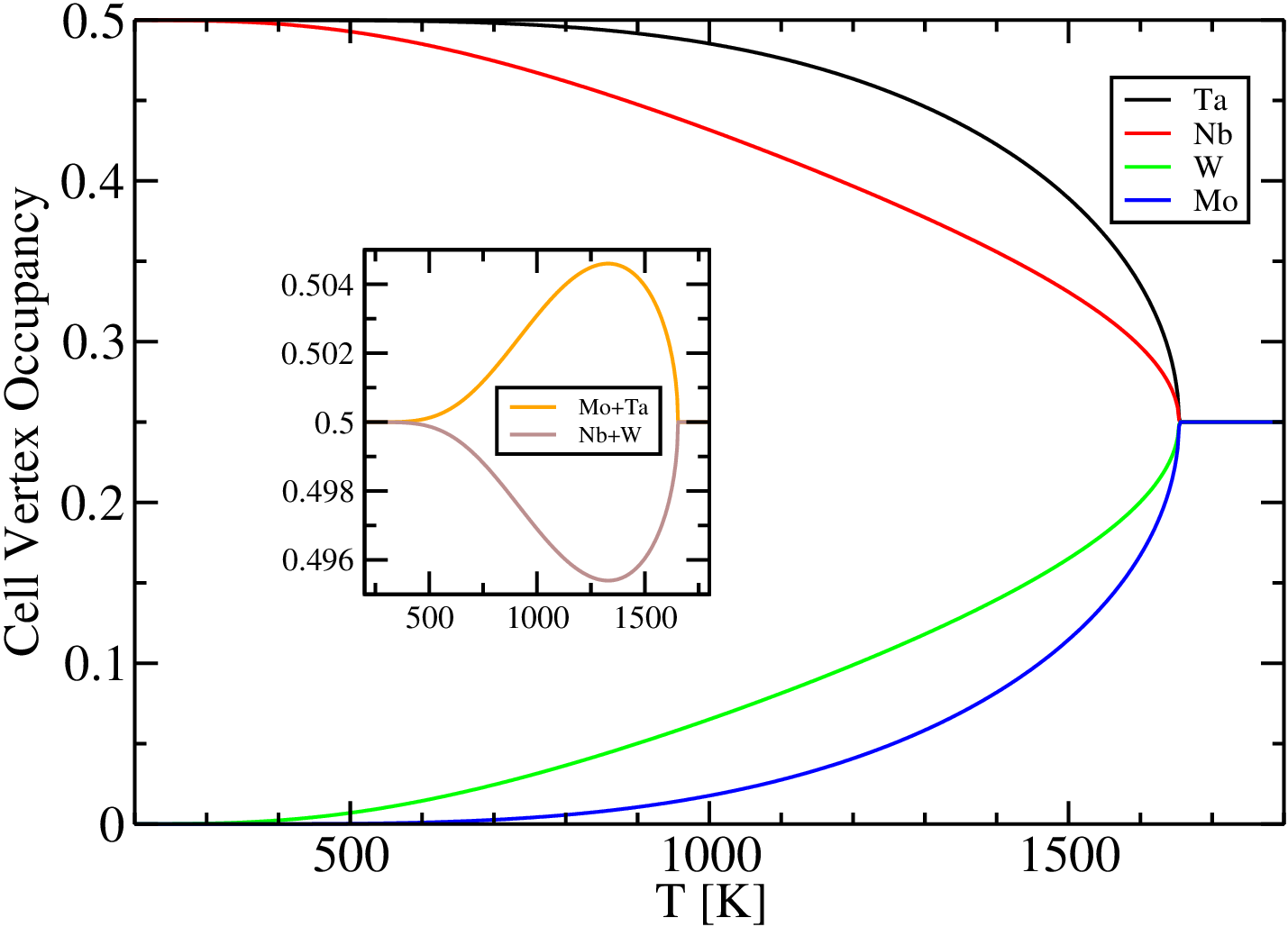}
	\caption{The vertex sublattice occupancy as a function of temperature.  See text for information on inset.}
	\label{fig:e_vs_T}
\end{figure}

Shown in Figure \ref{fig:e_vs_T} is the temperature evolution of the 
species concentration on cell vertices, obtained by minimizing the free 
energy at a given temperature over all independent variables $e_{\alpha'}$ 
subject to the bounds.  Monte Carlo simulations using the same enthalpy
model were subsequently performed to verify our mean field results, 
and they show excellent qualitative agreement.  Here the system achieves 
perfect sublattice occupancies for all species at low temperature, with 
Ta and Nb on the cell vertices and W and Mo on the cell centers.  This 
agrees with our pseudobinary model where group 5 (Ta,Nb) and group 6 
(W,Mo) form pseudobinary species.  As temperature rises, the system 
begins picking up some disorder, with Nb and W more strongly affected 
than Mo and Ta, finally reaching complete disorder at $T_{C}=1654K$.  Our
Monte Carlo simulations predict $T_{C}=1280K$, less than the mean field
value, as expected.  The temperature evolution of the enthalpy of this 
model (not shown) shows monotonic behavior up to the transition 
temperature, suggesting a second order transition.

\begin{table}
	\begin{tabular}{c||c|c|c||c}
	$\lambda$ & $e_{Ta}-1/4$ & $e_{Nb}-1/4$ & $e_{W}-1/4$ & $e_{Mo}-1/4$ \\
	\hline
	\hline
	-701 &  0.786 &  0.600 &  0.152 & -1.537 \\
	\hline
	   8 &  0.321 & -0.186 & -0.929 &  0.793 \\
	\hline
	  18 & -0.529 &  0.778 & -0.339 &  0.089 \\ 
	\hline
        \hline
        ${\bm \psi'}(T=0)$& 1/4 & 1/4 & -1/4   & -1/4
	\end{tabular}
	\caption{The eigenvalues and associated eigenvectors of $\theta$ using bond strengths given in Table \ref{tab:NNBondNum}.  $\lambda$ is in units of meV/atom.  The final column is the selected dependent species and is not part of the associated eigenvector; it is shown here to give physical intuition to how the particular mode is ordering the system.  Bottom row ${\bm \psi'}(T=0)$ is the superposition of eigenvectors ${\bm v}_i$ weighted by the low temperature normal coordinates $\chi_i(T=0)$.}
	\label{tab:eigenvecval}
\end{table}

The inset shows the temperature evolution of the quantities 
$e_{Ta}+e_{Mo}$ and $e_{Nb}+e_{W}$, both quantities remaining close to 
1/2 for all temperatures, explaining the mirror-image-like relation of  
Mo to Ta and Nb to W.  This arises from Mo-Ta's strong binding, pinning 
the configuration close to the boundary $e_{Ta} \approx e_{Nb} \lesssim \frac{1}{2}$, with $e_{W} \approx e_{Mo} 
\gtrsim 0$ at low temperatures.  Even at T $\approx$ 1000K, where 
$e_{Nb}$ and $e_{W}$ differ appreciably from their boundary values, 
$e_{Ta}$ and $e_{Mo}$ remain close to their extremes, requiring that 
$e_{Nb}+e_{W} = 1-(e_{Ta}+e_{Mo}) \approx \frac{1}{2}$.  By the time 
T reaches T\subscript{C}, $e_{\alpha}=\frac{1}{4}~\forall~\alpha$, 
so the values of $e_{Ta}+e_{Mo}$ and $e_{Nb}+e_{W}$ never deviate 
significantly from 1/2.

\begin{figure}
	\includegraphics[trim = 0mm 0mm 0mm 0mm, clip, width=0.75\textwidth]{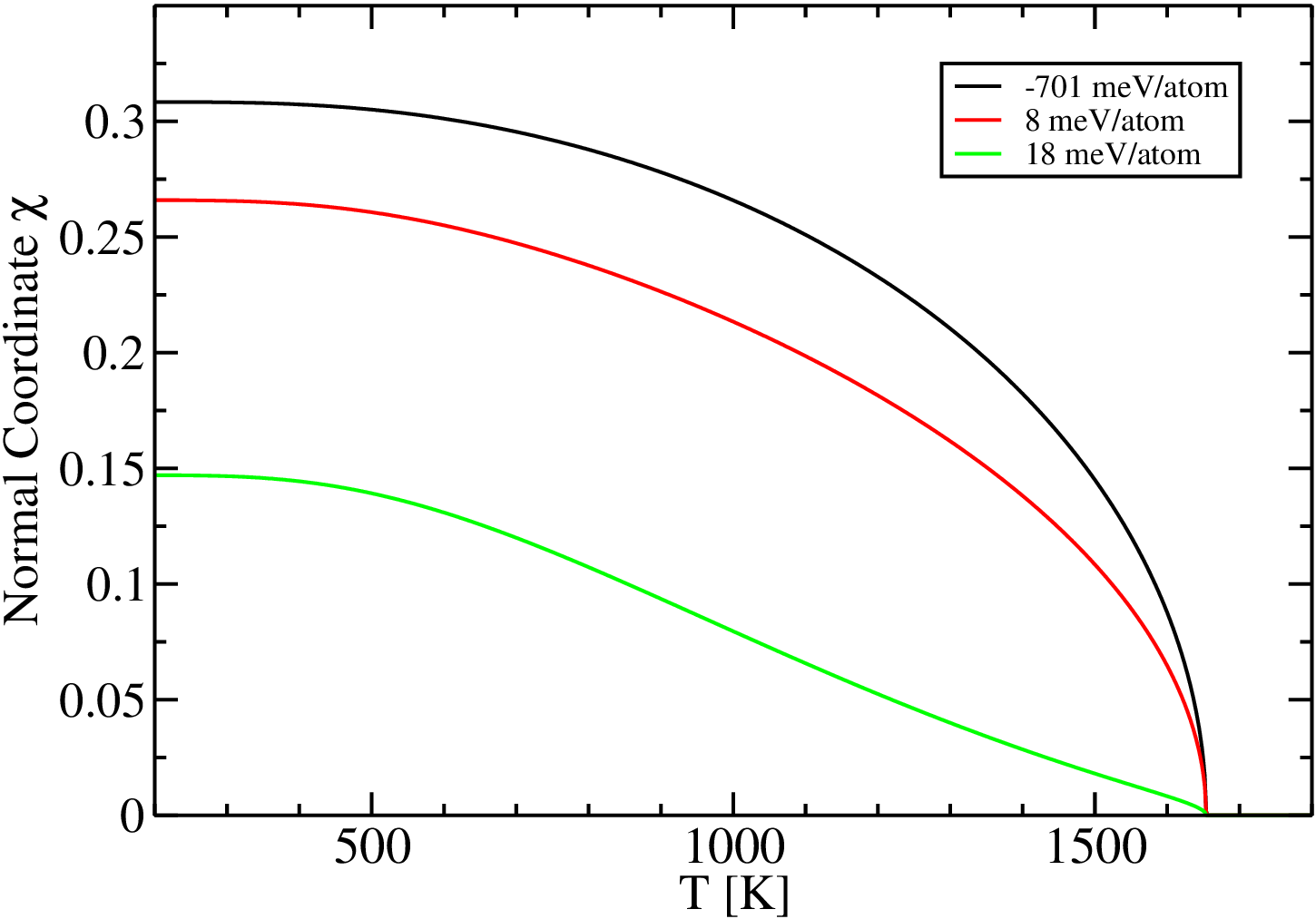}
	\caption{The normal coordinates of the system as function of temperature.}
	\label{fig:normalCoords_vs_T}
\end{figure}

Figure \ref{fig:normalCoords_vs_T} shows the temperature evolution of 
the normal coordinates of the system.  At low temperatures, all modes 
contribute to the thermodynamic equilibrium of the system, though the 
-701 meV/atom mode dominates the enthalpy of the system.  
All three normal coordinates vanish at $T_{C}=1654K$, yielding the 
disordered solution where entropy is maximized in configuration space.  
The square root singularity as T $\rightarrow$ T\subscript{C} matches 
the well-known mean field classical critical exponent 
$\beta = \frac{1}{2}$.

The system strongly prefers the -701 meV/atom mode at $T=0K$ due to 
its substantially low enthalpy.  This places Ta, Nb, and W on cell 
vertices and Mo on cell centers.  To maintain overall concentration 
1/4 for each species, the other two modes are needed to compensate.  
The 8 meV/atom mode replaces some Mo at cell centers with W, 
while further strengthening the ordering of Ta on cell vertices.  
Finally, the least favored 18 meV/atom mode is the only mode that 
orders Nb and W onto different sublattices, giving the observed 
disordering of Nb and W in Figure \ref{fig:e_vs_T}.  The large 
contribution of Mo-Ta ordering in the -701 meV/atom mode presents 
compelling evidence of ordering dominated by Mo-Ta bonding.  
 
\section*{Acknowledgements}
This work was supported in part by grant HDTRA1-11-1-0064. We thank 
Marek Mihalkovi\v{c} and Michael Gao for useful discussions.

\section{References}
\bibliographystyle{apsrev4-1}
\bibliography{wphbib}

\end{document}